\documentclass[runningheads]{llncs}
\usepackage{hyperref}
\usepackage{amsmath}
\usepackage{stmaryrd}
\usepackage{graphicx}
\usepackage{xspace}
\usepackage{xcolor}
\usepackage{cleveref}
\usepackage{wrapfig}
\usepackage{pgfplots}
\usepackage{proof}

\pgfplotsset{width=10cm,compat=1.9}

\usepackage{listings}

\definecolor{darkgreen}{rgb}{0,0.5,0} 

\newcommand{\semgus}{SemGuS\xspace}
\newcommand{\sygus}{SyGuS\xspace}
\newcommand{\chchead}[1]{\textrm{Sem}_{#1}}
\newcommand{\ksII}{\textsc{ks2}\xspace}

\usepackage[firstpage]{draftwatermark} 

\SetWatermarkAngle{0}
\SetWatermarkText{\raisebox{14.25cm}{%
\hspace{0.1cm}%
\href{https://doi.org/10.5281/zenodo.10947134}{\includegraphics{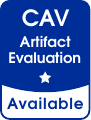}}%
\hspace{9cm}%
\includegraphics{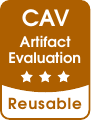}%
}}

\begin{document}
\title{The \semgus Toolkit}
%
%
\author{Keith J.C. Johnson\inst{1} \and
Andrew Reynolds\inst{2} \and
Thomas Reps\inst{1} \and
Loris D'Antoni\inst{1}}
%
%
\institute{University of Wisconsin--Madison \and
University of Iowa}
\maketitle              
\begin{abstract}
Semantics-Guided Synthesis (\semgus) is a programmable framework for defining synthesis problems in a domain- and solver-agnostic way. 
This paper presents the standardized \semgus format, together with
an open-source toolkit that provides
a parser, a  verifier, and enumerative \semgus solvers.
The paper also
describes an initial set of \semgus benchmarks, which form the basis for comparing \semgus solvers, and presents an evaluation of the baseline enumerative solvers.
\end{abstract}

\section{Introduction}
The field of program synthesis aims to create tools that can automatically create a program from a specification of desired behavior. Synthesis holds the promise of easing the burden on programmers (e.g., by finding solutions to tricky special cases automatically), and allowing non-programmers to create programs merely by indicating the outcome that they want the program to produce. 

While program synthesis has seen successes in many industrial applications~~\cite{polozov2015flashmeta,gulwani2012synthesis}, these successes have typically been achieved using domain-specific synthesizers that take advantage of the structure of the specific domain. 

To apply synthesis beyond specific domains, synthesis frameworks and tools should allow one to customize the search space and specifications of a synthesis problem in a programmable way that is agnostic of a specific domain or synthesis solver.
To address the problem of making synthesis ``programmable'', Kim et al.~\cite{kim2021semantics} proposed the \semgus framework, which enables one to specify synthesis problems in a solver-agnostic and domain-agnostic way~\cite{d2021programmable}.

The \semgus framework allows one to specify
an arbitrary synthesis problem by defining a programming language via (i) a grammar (the syntax), and (ii) a set of Constrained Horn Clauses (CHCs) (the semantics).
Once one has described the language, one can define synthesis problems over that language by providing a specification as a formula.
Solving the synthesis problem
means finding a program in the language that satisfies the specification.
Building solvers for general \semgus problems can be difficult due to the framework's flexibility~\cite{d2021programmable}.
%

This paper presents the \semgus toolkit, which provides an open-source implementation of the components needed for researchers to get started building \semgus solvers.
The toolkit consists of the following components.

\vspace{1mm}\noindent\textbf{\semgus Format 1.0:} The first standardized format for \semgus, which is built on top of the SMT-LIB and SyGuS formats~\cite{BarFT-RR-17,DBLP:journals/corr/abs-2312-06001}, thus making it expressible, extensible, modular, and easy to integrate with existing constraint solvers (e.g., to build \semgus verifiers).
We provide an open-source parser (\Cref{sec:format}).

\vspace{1mm}\noindent\textbf{Baseline Verifier and Solvers:} The flexibility of \semgus makes verifying whether a term is a solution to a \semgus problem undecidable.
Furthermore, because the semantics of the user-provided programming language is expressed declaratively using CHCs, it is even challenging to efficiently execute programs in the language. 
Our implementation provides a compiler that,
given a term $t$ in the user-specified language, can extract efficiently executable semantics for $t$ from the declarative one provided by the user, as well as
an incomplete SMT-based verifier that can construct constraints for checking whether $t$ matches a specification $\varphi$.
We also provide implementations of top-down and bottom-up example-based enumerative solvers that are integrated with these verifiers and can thus produce solutions to \semgus problems (\Cref{sec:baseline-tool}). 

\vspace{1mm}\noindent\textbf{Benchmarks:} We provide 431 \semgus benchmarks from different domains. Our solvers can only solve 161/431 benchmarks, and we hope this toolkit will energize the community to build solvers for the remaining challenging problems and to provide additional benchmarks (\Cref{sec:benchmarks}).

\section{The \semgus Format 1.0}
\label{sec:format}

We refer the reader to the original \semgus paper~\cite{kim2021semantics} for a more formal definition of the \semgus framework, but in this section we show how each component is expressed in our proposed standard format.
%
The \semgus parser (\url{https://github.com/semgus-git/Semgus-Parser}) can translate the textual \semgus format into two intermediate representations: a JSON format and a declarative S-expression format, which is then used by solvers and other tools.

Figures~\ref{fig:language-definition} and~\ref{fig:constraints} give an example specification of a \semgus problem, which we describe in detail in this section.
%
%
In this example,
the goal is to synthesize an imperative program (with loops) that multiplies two numbers through iterative addition.
We choose this example because it illustrates how \semgus can describe synthesis problems involving complex programming constructs and is thus strictly more expressive than limited synthesis frameworks, such as \sygus~\cite{sygus}.



\vspace{1mm}\noindent\textbf{Term Universe.}
\semgus problems define a universe of terms with a modified SMT datatype declaration
using the command \texttt{declare-term-types} (lines~\ref{line:syntax-start}--\ref{line:syntax-end} of Figure~\ref{fig:language-definition}).
This command defines the syntax of the programming language over which
one can specify synthesis problems.
The term universe $L$ is intentionally separated from the sub-universe (defined by a grammar) from which the answer is to be synthesized, and from the constraints on the answer (Figure~\ref{fig:constraints}).
The user defines $L$ and its semantics once and for all, and can reuse those definitions for different synthesis problems.
This separation enables both \textit{(i)} building specialized \semgus solvers for important languages (e.g., $L = \textrm{SQL}$), and \textit{(ii)} instantiating more restricted synthesis problems by confining the search space to just the terms generated by a grammar.

\begin{figure}
    \centering
\begin{lstlisting}
;; Nonterminals /*@ \label{line:syntax-start} @*/
(declare-term-types ((F 0) (S 0) (E 0) (B 0))
;; Term Universe (i.e., language syntax)
((($function S E))                             ;; F
 (($x<- E) ($y<- E) ($r<- E) 
  ($noop) ($seq S S) ($while B S))             ;; S
 (($r) ($0) ($1) ($x) ($y) ($+ E E) ($- E E))  ;; E
 (($< E E))))                                  ;; B   /*@ \label{line:syntax-end} @*/
;; Constrained Horn Clauses (i.e., language semantics) /*@ \label{line:sem-start} @*/
(define-funs-rec
 ;; Types of semantic relations   /*@ \label{line:sem-rel-start} @*/
 ((F.Sem ((t F) (x Int) (y Int) (ret Int)) Bool)
  (S.Sem ((t S) (xi Int) (yi Int) (ri Int) (xo Int) /*@ \label{line:s-rel} @*/
          (yo Int) (ro Int)) Bool)
  (E.Sem ((t E) (xi Int) (yi Int) (ri Int) (out Int)) Bool)
  (B.Sem ((t B) (xi Int) (yi Int) (ri Int) (out Bool)) Bool))/*@ \label{line:sem-rel-end}@*/
 ;; CHCs defining semantic relations  /*@ \label{line:CHCs-start} @*/
 (;; Semantics of functions
  (! (match t (^...^)) :input (x y) :output (ret))    
  ;; Semantics of statements
  (! (match t
     ^...more S productions...^
     (($noop)         ;; Noop statement
      (and (= xi xo) (= yi yo) (= ri ro)))
     (($seq t1 t2)    ;; Sequential composition 
      (exists ((x1 Int) (y1 Int) (r1 Int))
        (and (S.Sem t1 xi yi ri x1 y1 r1) 
             (S.Sem t2 x1 y1 r1 xo yo ro))))
     (($while tb ts)  ;; While statement /*@ \label{line:while-start} @*/
      (exists ((b Bool) (x1 Int) (y1 Int) (r1 Int))
        (and (B.Sem tb xi yi ri b) ;; While-true
             (= b true)
             (S.Sem ts xi yi ri x1 y1 r1)
             (S.Sem t x1 y1 r1 xo yo ro)))
      (exists ((b Bool))
        (and (B.Sem tb xi yi ri b) ;; While-false
             (= b false)
             (= xo xi) (= yo yi) (= ro ri)))))) /*@ \label{line:while-end} @*/
  :input (xi yi ri) :output (xo yo ro))     
  ;; Semantics of integer expressions
  (! (match t (^...^)) :input (xi yi ri) :output (out))
  ;; Semantics of Boolean expressions
  (! (match t (^...^)) :input (xi yi ri) :output (out))))/*@ \label{line:sem-end} @*/
\end{lstlisting}
    \caption{Definition of a programming language (i.e., a set of programs) in the \semgus format.
    The syntax of terms is given in lines~\ref{line:syntax-start}--\ref{line:syntax-end}, and their semantics is given in lines~\ref{line:sem-start}--\ref{line:sem-end}.
    The {\color{gray}{\texttt{gray text}}} denotes parts that have been omitted for brevity. 
    }
    \label{fig:language-definition}
\end{figure}

\vspace{1mm}\noindent\textbf{Semantics as CHCs.} 
The semantics of our term language is given by the SMT-LIB command \texttt{define-funs-rec} (lines~\ref{line:sem-start}--\ref{line:sem-end}).
In a nutshell, this command defines a set of Constrained Horn Clauses (CHCs) inductively over terms in the universe.
A CHC is a first-order formula of the form:
\[\forall \bar{x}_1, \dots, \bar{x}_n, \bar{x}.~\phi \land R_1(\bar{x}_1) \land \dots \land R_n(\bar{x}_n) \Rightarrow H(\bar{x})\]
where $R_1, \dots, R_n$ and $H$ are uninterpreted relations, $\bar{x}_1, \dots, \bar{x}_n$ and $\bar{x}$ are (vectors/tuples of) variables, and $\phi$ is a quantifier-free constraint over the variables within some first-order theory. 
In the specification, one provides the names and types of the semantic relations used in the semantic definitions
(lines~\ref{line:sem-rel-start}--\ref{line:sem-rel-end}), and then CHCs that define such relations (lines~\ref{line:CHCs-start}--\ref{line:sem-end}).
To better align with the fact that CHCs are used to define the semantics of programs inductively (i.e., as an interpreter), in the \semgus format we encode CHCs as a set of mutually-recursive SMT functions, taking the term to be evaluated, input variables, and output variables as arguments, and returning a Boolean.
A function is provided for every non-terminal, and \texttt{match} statements are used to
dispatch on the term constructors for which one is defining the semantics.
The match statement must match on all productions for the given term type (i.e., the corresponding nonterminal). 
The match statement can also be annotated with which variables are inputs and outputs in the specific semantics (note that some semantics, e.g., a term-rewriting system, do not necessarily have inputs and outputs).
Each match on a production starts with an optional \texttt{exists} block, which specifies auxiliary variables, followed by the CHC body as a conjunction. 
Some productions, such as the \texttt{$while$} production in Figure~\ref{fig:language-definition} (lines~\ref{line:while-start}--\ref{line:while-end}), have two associated CHCs.
For example, the \texttt{While-false} CHC can be logically written as
\[
\infer[\texttt{While-false}]{S.Sem((\texttt{\$while}~tb~ts), xi, yi, ri,xo,yo,ro) }{
  B.Sem(tb,xi,yi,ri,b) & b=\textit{false} \wedge xo=xi \wedge yo=yi \wedge ro=ri 
}
\]
The signature at the bottom of the CHC---i.e., the particular variables names used in this relation instance---is the one defined in Figure~\ref{fig:language-definition} (line~\ref{line:s-rel}).

As discussed in the original \semgus paper~\cite{kim2021semantics}, many synthesizers have achieved scalability by exploiting alternative semantics that either underapproximate
the actual semantics of the programming language (to speed up evaluation and enable constraint solving) or overapproximate it
(which sometimes makes it possible to prune the search space of programs).
While such previous work ``hardcodes'' and takes advantage of such semantics in the solver itself,
\semgus allows one to write such semantics directly in the \semgus file.
In fact, there is no limit on how many semantic relations one can define in a \semgus file.
For example, one might define a semantic relation that associates costs to programs (but does not evaluate programs) and a semantic relation that captures program evaluation.
The specification can then require finding a program that (i) performs a computation correctly, and (ii) has a cost that is less than a specific constant.
The ability of \semgus to describe multiple semantics enables reusable solving techniques and interoperability between solvers.

It should be noted that when one defines multiple semantics, the burden of showing that they are properly related (e.g., 
that an abstract semantics is related to the concrete one by a Galois connection \cite{DBLP:conf/popl/CousotC77}) 
is in the hands of the user.
Doing so automatically is a research direction enabled by the \semgus format.



\vspace{1mm}\noindent\textbf{\texttt{synth-fun} Command.}
\semgus uses the same syntax as in \sygus to declare what type of term we are interested in synthesizing (Figure~\ref{fig:constraints}). 
Unlike \sygus, a solution to a \semgus problem is a term in the provided syntax, as opposed to a function in an SMT theory. 
For instance, the command \texttt{(synth-fun mul () F)} in Figure~\ref{fig:constraints} (line~\ref{line:synthfun}) asks for
a term named \texttt{mul}, rooted at the non-terminal \texttt{F}.
This command can optionally take a grammar (the second argument) to further restrict the search space, using the same format for grammars as in the \sygus format~\cite{DBLP:journals/corr/abs-2312-06001}.
For example, to synthesize programs that have the fewest number of while-loops~\cite{qsygus}, one might first solve the \semgus problem discussed in this section and obtain a program with one loop, and then create a new \semgus problem where the grammar is restricted to disallow loops.
The two problems will share the same language definition despite having different grammars.

\vspace{1mm}\noindent\textbf{Specification.} Specification constraints for \semgus  problems are stated using
\begin{wrapfigure}{l}{0.51\textwidth}
\vspace{-7mm}
    \centering
    \begin{lstlisting}
;; Function to synthesize    
(synth-fun mul () F)  /*@ \label{line:synthfun} @*/
;; Constraints for examples
(constraint (F.Sem mul 0 0 0))/*@ \label{line:constraints-start} @*/
(constraint (F.Sem mul 1 1 1))
(constraint (F.Sem mul 2 2 4))
(constraint (F.Sem mul 3 3 9))
(constraint (F.Sem mul 5 3 15))
(constraint (F.Sem mul 3 4 12))/*@\label{line:constraints-end}@*/
;; Perform synthesis
(check-synth)  
    \end{lstlisting}
\vspace{-4mm}    
    \caption{
    Constraints for a few example input/output pairs, used to synthesize a function \texttt{mul} that behaves like multiplication.
    }
\vspace{-5mm}        
    \label{fig:constraints}
\end{wrapfigure}
 SMT expressions involving the root CHC for the term to be synthesized.
The typical form for input/output examples is shown in Figure~\ref{fig:constraints} (lines~\ref{line:constraints-start}--\ref{line:constraints-end}). 
Note that in \semgus, 
constraints are specified as relations and not functions (as in \sygus).
Relations allow modeling nondeterminism or nonterminating semantics---e.g., one can state that, for the specific input pair (5,3), the answer is a positive value if the program terminates:
\texttt{(constraint (forall ((x Int)) (=> (F.Sem mul 5 3 x) (>= x 0))))}.


\vspace{1mm}\noindent\textbf{Synthesis Command.} 
The \texttt{check-synth} command instructs the solver to solve the problem and produce an SMT term.
The following term is a solution to the example presented in this section.
\begin{lstlisting}
((define-fun mul () F ($function
        ($while ($< $0 $y)              ;; while (0<y)
            ($seq ($y<- ($- $y $1))     ;;     y <- y-1
                  ($r<- ($+ $r $x))))   ;;     r <- r+x
        $r)))                           ;; return r
\end{lstlisting}

\vspace{1mm}\noindent\textbf{Relationship between \semgus and \sygus.}
%
Every \sygus problem can be automatically converted to an equivalent \semgus problem, 
and our parser implements this transformation.
The only technical detail of interest is that \sygus synthesizes \textit{function SMT terms}, whereas \semgus synthesizes terms in
a term universe that is interpreted using a \textit{relational semantics}.
For example, if the predicate $\varphi$ of the \sygus specification contains invocations of  the function  $g$ to be synthesized, e.g., $\varphi(g(i_1), \ldots, g(i_n))$, we can create the new \semgus specification as $\exists o_1,\ldots,o_n.\; \varphi(o_1,\ldots,o_n) \wedge \chchead{G}(g, i_1, o_1) \wedge \ldots \wedge \chchead{G}(g, i_n, o_n)$.




Because \semgus is more expressive than \sygus, not every \semgus problem can be converted to an equivalent \sygus problem.
In general, it is
undecidable to check when such a translation is possible, because \semgus is Turing complete.
We have implemented a sound (but incomplete) translation of a limited fragment of statically detectable \semgus problems into \sygus.
The fragment essentially captures when the \sygus-to-\semgus translation can be inverted.



\section{A Baseline \semgus Solver}
\label{sec:baseline-tool}
In this section, we present \ksII, a toolkit for researchers to build \semgus solvers.
\ksII implements techniques for (efficiently) verifying whether a candidate solution meets the specification (\Cref{sec:verifier}).
\ksII also contains implementations of bottom-up and top-down enumerative synthesizers (\Cref{sec:solvers}).
\ksII is written in Common Lisp,
which makes it easy to compile code generated at synthesis-time for speeding up evaluation of candidate solutions.
In addition, \ksII is implemented modularly, so new solvers and features 
can be easily added as plugins.

\subsection{Verifying Candidate Solutions}
\label{sec:verifier}

When building synthesizers, one wants two types of verifiers: one that can quickly tell if a candidate solution is correct on a finite set of input examples $E$, and one that can (less quickly) tell if a solution is correct on all inputs, and thus satisfies a logical specification.
When the latter verifier finds a violation of the specification, it will typically produce a new input example $e$ that can be added to the set $E$ to restart synthesis with a fresh set of examples.
These two verifiers together form the basis of the counterexample-guided synthesis algorithm.
For \semgus, building either of these verifiers is generally undecidable as one may have to deal with an arbitrarily powerful programming language.

In this section, we present two sound (but incomplete) implementations of such verifiers.
These implementations are not the only verifier implementations that can be built for
SemGuS, but just two that were successful in meeting our needs. Building other verifier implementations based on other technologies, such as bounded model checkers, symbolic execution, and logic programming, is an interesting future research direction.

\vspace{1mm}\noindent\textbf{Building Executable Semantics from CHCs.}
To tell quickly whether a candidate program is correct on a given input, one needs to ``run'' the program on the input according to the semantics.
To do so efficiently is
nontrivial
because the semantics of
a candidate program
is expressed declaratively using CHCs.
\ksII first ``operationalizes'' the semantics given by CHCs into executable blocks,
which are then compiled.
In general, not all CHCs can be transformed into executable code (for example, non-deterministic CHCs that can map one input to different outputs);
therefore, \ksII supports only a fragment of CHCs that is practically useful (all benchmarks discussed in \Cref{sec:benchmarks} fall into this fragment).
%

We illustrate the compilation to native code using the following (recursive) CHC corresponding to the \texttt{While-true} case in \Cref{fig:language-definition}:
\[
\chchead{S}(t(t_b, t_s), i, o) \impliedby \chchead{B}(t_b, i, b) \wedge b \wedge \chchead{S}(t_s, i, o') \wedge \chchead{S}(t, o', o) 
\]
\ksII requires each {position} in each relation to be annotated as an input or output variable.\footnote{
  Automatically inferring such annotations (``mode inference'') is a classical analysis problem in logic programming \cite[\S10.2.2]{DBLP:journals/jlp/CousotC92}.
  Automatically supplying annotations is an interesting research direction for \semgus.
}
In the example, the first position of $\chchead{S}$ and $\chchead{B}$ is the term being executed, which is always assumed to be an input;
the second position is the input on which the term should be executed, and the last position is the output.
To operationalize the CHC, \ksII performs the following steps
to identify an evaluation order:
it analyzes each relation instance in the body of the CHC, and performs a dataflow analysis to
determine an order in which the blocks can be executed.
This step is done
by building a dataflow graph and then performing a topological sort to identify an order in which each relation can be computed.
In the given example,
one possible order is to first evaluate $\chchead{B}(t_b, i, b)$ because $i$ is readily available, then determine whether $b$ is true, then evaluate $\chchead{S}(t_s, i, o')$, 
and finally evaluate $\chchead{S}(t, o', o)$ ($o'$ depends on one of the previous relations). 
Our implementation of this transformation has some basic requirements.
First, no two relations can output the same variable---otherwise one cannot resolve which instance to use.
Second, every input variable $i'$ to a relation is either the output of another relation or appears in the first-order formula of the CHC in the form $i'=f(\cdot)$---i.e., the value of $i'$ can be computed without having to call an SMT solver.

At this point, we generate code for each block. Child CHCs turn into function calls, guards into conditional statements, and value productions into assignments. This generated code is then compiled and 
turned into an executable function that implements the CHC's semantics.
To execute a program on an input/output example, the top-level semantic function is called with the input state and the child's semantic functions, and the program returns the output state. This output state can be checked against the output example.

This implementation of an efficiently executable semantics is one of the main contributions in \ksII.
Identifying
additional
ways to compile the logical semantics into
an efficiently-executable one
is an interesting research direction that can benefit from techniques in compiler design and logic programming.

\vspace{1mm}\noindent\textbf{A Simple Incomplete Verifier for Logical Specifications.}
The declarative nature of \semgus enables a simple way of building a verifier that can check a program against a specification and return a counterexample.
As we argued, verification for \semgus is undecidable, but the declarative nature of \semgus allows us to build a simple, but incomplete, procedure for verifying some candidates in \semgus solutions.
Given a concrete term $t$ (i.e., the program we are trying to verify), our verifier performs
a pre-order traversal of $t$ and emits a potentially recursive SMT function for each node that corresponds to the CHC (or CHCs) for that node.
Because $t$ is a concrete term, each child term in the CHC can also be replaced by 
the concrete function implementing it. For example, the program \texttt{(\$+ \$x \$1)}
would be verified by emitting three SMT functions for \texttt{\$+}, \texttt{\$x}, and \texttt{\$1}.
The function for \texttt{\$+} will call the ones for \texttt{\$x} and \texttt{\$1} to perform the evaluation.
Operators like \texttt{\$while} require recursive function calls.
The specification can then be used to define constraints over the root node and verified with an
SMT solver. 
In the case of recursive semantics, the SMT functions will potentially be mutually recursive, thus relying on undecidable theories for which current SMT solvers struggle in practice.
This verifier, while incomplete, is ``good enough'' for  many of our current benchmarks, and extending SMT solvers or our verifiers to better handle such cases is a challenging research question.

\subsection{Baseline Enumerative solvers}
\label{sec:solvers}

\ksII implements standard basic top-down and bottom-up enumeration algorithms as described in the literature~\cite{PGL-010}.
Because we have a logical verifier that enables counterexample-guided inductive synthesis (CEGIS), our enumeration algorithms only check correctness on a set of examples.

For top-down enumeration, candidate programs are enumerated using a priority queue of potentially partial programs (i.e., with holes).
At each iteration a program is extracted from the queue:
if it has no holes, it is verified against the examples; otherwise, all the programs that can be obtained by expanding the leftmost hole with all possible child productions are added to the queue. 
The standard optimization for top-down enumerators is to check partial programs against the specification and prune them if possible.
Existing optimizations are domain-specific and identifying ways to extend them to \semgus problems is an open research question, which we hope this toolkit will help researchers work on.

For bottom-up enumeration, subterms of increasing size (or height) are enumerated and added to a program bank where they are grouped by size (or height).
Enumeration of programs of a certain size or height happens lazily; they are verified and pushed
into the bank of programs one at a time.
A typical optimization used in a bottom-up enumeration is to use some form of equivalence-checking to deduplicate enumerated programs with the same behavior. 
One popular technique, observational equivalence, executes each enumerated program on the input example states and prunes a program if a previously enumerated program returns the same output state. However, because \semgus supports imperative semantics, the possible input states for a sub-program 
are not necessarily the same as the top-level-program's input states (i.e., variable values change throughout the program execution), and thus there is not an easy way to perform an observational-equivalence check.
The development of an appropriate pruning technique for a bottom-up \semgus enumerator is an open research question, which we hope this toolkit will help researchers work on, for example by building on approaches such as equality saturation~\cite{egg} and lifting interpretation to sets of programs~\cite{10.1145/3632894}.



\subsection{Extensibility}
\label{sec:extensibility}
\ksII can be extended by instantiating various interfaces with modules. For example, one might want to add a module that implements a technique for pruning enumerated programs with the bottom-up enumeration. To add this technique, the module would implement the \texttt{add-to-bank} interface, which is responsible for adding freshly enumerated programs to the bank of enumerated programs, and simply decline to add programs that the module can prune. In code, this implementation might look like:
\begin{lstlisting}
(defmethod add-to-bank :around ((ext prune) bank prog metric)
  "Adds the program PROG to BANK unless it should be pruned"
  (unless (%should-prune prog) (call-next-method)))
\end{lstlisting}
where \texttt{prune} is the module class and \texttt{\%should-prune} implements the
predicate for whether or not a program should be pruned.
At this time, among others, we have interfaces for adding solvers, adding verifiers, and inspecting and updating the \semgus problem. We will continue to add more interfaces as the need arises; the most up-to-date documentation is available with \ksII and its supporting libraries.

Outside of \ksII, the \semgus Parser is available as a standalone tool for parsing \semgus problems into JSON, as well as a .NET library for direct integration into solvers. We expect these parsing tools to lower the barrier to entry for building new \semgus tooling.

\section{Benchmarks and Performance of Baseline Solvers}
\label{sec:benchmarks}

We present an initial set of \semgus benchmarks and evaluate the performance of our baseline solvers on such benchmarks.

\vspace{1mm}\noindent\textbf{Benchmarks.}
The ability of \semgus to represent synthesis problems from disparate domains in the same solver-agnostic format is one of its key distinguishing features.
We have created 431 \semgus benchmarks,
consisting of synthesis problems from a variety of domains.
\begin{description}
    \item[Sample domains:] 17 benchmarks of easy synthesis problems (10 for imperative programs with loops, 3 for SMT datatypes, and 4 integer-arithmetic benchmarks). These benchmarks are designed to help researchers build \semgus solvers and are basic test of a solver's support of various features of the \semgus format. They contain between 1 and 6 input/output examples each.
    
    \item[Regular expressions:] 72 benchmarks for synthesizing regular expressions, which include problems from the original \semgus paper~\cite{kim2021semantics}, from the tool AlphaRegex~\cite{alpharegex}, and CSV formatting problems. These benchmarks have between 2 and 244 input/output examples each. Benchmarks in this category may use two different semantics of regular expressions: one based on Boolean matrices and one based on SMT terms for the theory of regular expressions.
    \item[Boolean formulas:] 88 benchmarks for synthesizing Boolean formulas, including DNF (32), CNF (33), and cube (23) formulas.
    Each benchmark has between 4 and 128 input/output examples.
    \item[Bitvectors:] We provide 100 benchmarks over imperative loop-free bitvector programs~\cite{10.1145/1993498.1993506}.
    In our adaptation of the existing benchmarks, we consider different bitvector semantics (e.g., one where bitvectors restart at 0 on overflow, and one where the values remain at INT\_MIN or INT\_MAX).
    The ability to customize programs semantics is a key feature of \semgus.
    These benchmarks use logical specifications instead of input-output examples.
    \item[Messy:] 154 benchmarks (15 bitvector, 18 imperative, 121 unrealizable \sygus and imperative) from the original \semgus paper.  
\end{description}

We expect this set to be extended. New benchmarks may be submitted to the Semgus-Benchmarks GitHub repository via pull requests. All submissions are automatically checked for proper syntax and manually reviewed by maintainers for appropriateness before being included.

\begin{table}[t]
    \centering
\setlength{\tabcolsep}{4pt}    
\caption{Solved benchmarks by category. 
}
    \label{tab:benchmark-numbers}    
    \begin{tabular}{c|c|ccc|c}
        Domain & Total & TopDown & BottomUp(H) & BottomUp(S) &  Virtual Best \\ \hline
        Sample Domains & 17 & 14 & 11 & 13  & 15 \\        
        Regular Expressions & 72 & 52 & 8 & 45 & 54 \\                
        Boolean & 88 & 45 & 47 & 46 &  49 \\
        Bitvectors & 100 & 38 &27 & 36  & 43 \\        
        Messy & 154 & 0 & 0 & 0 & 0 \\
        \hline
        Total & 431 & 149 & 93 & 140 & 161 \\
    \end{tabular}    
\end{table}

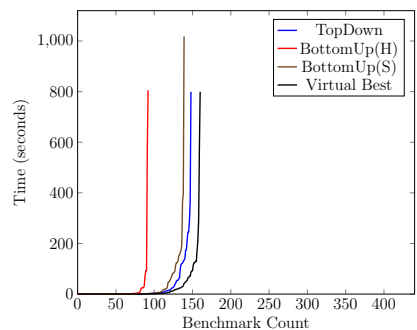
\begin{wrapfigure}{r}{0.45\textwidth}
    \centering
     \scalebox{0.43}{
    \begin{tikzpicture}
    \begin{axis}[        
        ylabel={Time (seconds)},
        xlabel={Benchmark Count},
        xmin=0,
        xmax=440,
        ymin=0,
        width=12cm,
        tick label style={font=\Large},
        label style={font=\Large},        
        legend style={font=\Large},
        legend pos=north east,
        no markers]
       \addplot+[sharp plot,mark=none,very thick] table [x index=0,y index=3] {figs/data/runtime.dat};  
       \addplot+[sharp plot,mark=none,very thick] table [x index=0,y index=1] {figs/data/runtime.dat};
       \addplot+[sharp plot,mark=none,very thick] table [x index=0,y index=2] {figs/data/runtime.dat};     
       \addplot+[sharp plot,mark=none,very thick] table [x index=0,y index=4] {figs/data/runtime.dat};
       \legend{TopDown, BottomUp(H),BottomUp(S), Virtual Best}
    \end{axis}
\end{tikzpicture}
    }
    \vspace{-6mm}
    \caption{Cactus plot of runtime.
    (Lower and to the right is better.)
    }
    \vspace{-6mm}
    \label{fig:runtime-cactus}
\end{wrapfigure}

\vspace{1mm}\noindent\textbf{Performance of Baseline Solvers.}
Benchmark results for our top-down and bottom-up enumerators by height (H) and size (S) are 
shown \Cref{tab:benchmark-numbers} as a summary solved instances and \Cref{fig:runtime-cactus} as a cactus plot illustrating the time taken to solve the benchmarks.
All experiments are run on a cluster~\cite{https://doi.org/10.21231/gnt1-hw21}, with each
node having an AMD EPYC 7763 64-Core Processor, of which we requested two cores and
12 GiB of RAM. 
We set a timeout value of 2000s and memory limit of 8 GiB. We run each experiment $5$ times and report the median of these runs.

For Sample Domains, the solvers performed similarly and cumulatively solved 15/17 benchmarks (Virtual Best).
Top-down enumeration is a clear winner for Regular Expressions, with height-based bottom-up enumerator performing poorly because the solutions are typically narrow-but-tall. 
All solvers performed about equivalently on the Boolean benchmarks, although each solver solves a slightly different subset of the problems.
For Bitvectors, the bottom-up height-based solver underperformed because these benchmarks have grammars with many productions per non-terminal, thus producing many programs at each height.
However, size-based bottom-up enumeration could solve 5 problems that the top-down enumerator could not solve, and the top-down enumerator solved 7 that the bottom-up, size-based enumerator could not solve.
Note that the Bitvector benchmarks have relational specifications and were solved with CEGIS, but for 19 benchmarks, the verifier failed to check a candidate program or generate counterexamples for at least one solver. For the 43 solved benchmarks, the verifier generated between 1 and 10 counterexamples (average 4.5), in less than 150ms each (average 30ms). The remaining 38 benchmarks generated up to 12 counterexamples before exceeding the timeout or memory limit.
Our solver could not solve any Messy benchmarks: most are unrealizable (i.e., they have no solution) or use specifications that are hard to verify using  \ksII's SMT-based verifier. 
The Messy solver is particularly good at proving problems unrealizable, but it has not been ported to the \semgus format and we cannot include it in our baseline.

In terms of enumeration throughput (enumerated programs per second), our solvers perform similarly, and they can enumerate up to 150,000 programs per second (average 33,000) for benchmarks for which verification is quick.
The advantage of building and using executable semantics is obvious: if the logical verifier is instead called on each candidate, the throughput drops to at most 800 programs per second (average 175). On the benchmarks where solving with executable semantics and the logical verifier are both supported, the use of the executable semantics is on average 220 times faster than the logical verifier (geomean).

These results provide a baseline rate for future 
\semgus solvers to be compared against; the advantage of simple enumerators is their raw speed.



\section{Related Work}

The syntax-guided-synthesis paradigm~\cite{sygus} has been successfully used in many applications, including invariant synthesis~\cite{DBLP:conf/pldi/MiltnerPMW20,DBLP:conf/cav/FedyukovichPMG19}, and synthesis of rewrite rules and invertibility conditions~\cite{DBLP:conf/sat/NotzliRBNPBT19,DBLP:conf/cav/NiemetzPRBT18}. 
Several efficient solvers are available for this format~\cite{DBLP:conf/cav/ReynoldsBNBT19,DBLP:conf/tacas/AlurRU17,DBLP:conf/pldi/HuangQSW20}.
This effort has inspired several domain-specific extensions for domains that cannot be captured by standard SMT-LIB theories~\cite{DBLP:journals/corr/abs-2312-06001}.
In contrast, this work develops a general framework for which these extensions can be expressed in a uniform way.
Moreover, \semgus allows one to define synthesis problems---e.g. for imperative programs---that cannot be captured in a natural way by an SMT theory.
The syntax-guided-synthesis paradigm has been extended to signatures with oracles~\cite{DBLP:conf/icse/JhaGST10,DBLP:conf/vmcai/PolgreenRS22}, or symbols whose semantics are given by user-provided binaries.
In contrast, in SemGuS, the semantics of all symbols are fully expressed in the problem description.

\section*{Acknowledgements}
The authors would like to thank Jinwoo Kim, for initial discussions about the \semgus format; Wiley Corning, Rahul Krishnan, and Shaan Nagy, for code contributions to the \semgus parser; Evan Geng, Jiangyi Liu, and Charlie Murphy for finding and reporting bugs; and, in addition to everyone previously listed, Kanghee Park, Anvay Grover, and all future contributors for providing \semgus benchmarks.

Supported, in part, by
a Microsoft Faculty Fellowship;
a gift from Rajiv and Ritu Batra;
and NSF under grants CCF-\{1750965, 1918211, 2023222, 2211968, 2212558\}.
Any opinions, findings, and conclusions or recommendations expressed in this publication are those of the authors, and \
do not necessarily reflect the views of the sponsoring entities.

%
%
%
\bibliographystyle{splncs04}
\bibliography{refs}

\begin{thebibliography}{10}
\providecommand{\url}[1]{\texttt{#1}}
\providecommand{\urlprefix}{URL }
\providecommand{\doi}[1]{https://doi.org/#1}

\bibitem{sygus}
Alur, R., Bodik, R., Juniwal, G., Martin, M.M.K., Raghothaman, M., Seshia,
  S.A., Singh, R., Solar-Lezama, A., Torlak, E., Udupa, A.: Syntax-guided
  synthesis. In: 2013 Formal Methods in Computer-Aided Design. pp.~1--8 (2013).
  \doi{10.1109/FMCAD.2013.6679385}

\bibitem{DBLP:conf/tacas/AlurRU17}
Alur, R., Radhakrishna, A., Udupa, A.: Scaling enumerative program synthesis
  via divide and conquer. In: Legay, A., Margaria, T. (eds.) Tools and
  Algorithms for the Construction and Analysis of Systems - 23rd International
  Conference, {TACAS} 2017, Held as Part of the European Joint Conferences on
  Theory and Practice of Software, {ETAPS} 2017, Uppsala, Sweden, April 22-29,
  2017, Proceedings, Part {I}. Lecture Notes in Computer Science, vol. 10205,
  pp. 319--336 (2017). \doi{10.1007/978-3-662-54577-5\_18},
  \url{https://doi.org/10.1007/978-3-662-54577-5\_18}

\bibitem{BarFT-RR-17}
Barrett, C., Fontaine, P., Tinelli, C.: {The SMT-LIB Standard: Version 2.6}.
  Tech. rep., Department of Computer Science, The University of Iowa (2017),
  available at \url{www.SMT-LIB.org}

\bibitem{https://doi.org/10.21231/gnt1-hw21}
{Center for High Throughput Computing}: Center for high throughput computing
  (2006). \doi{10.21231/GNT1-HW21}, \url{https://chtc.cs.wisc.edu/}

\bibitem{DBLP:conf/popl/CousotC77}
Cousot, P., Cousot, R.: Abstract interpretation: {A} unified lattice model for
  static analysis of programs by construction or approximation of fixpoints.
  In: Graham, R.M., Harrison, M.A., Sethi, R. (eds.) Conference Record of the
  Fourth {ACM} Symposium on Principles of Programming Languages, Los Angeles,
  California, USA, January 1977. pp. 238--252. {ACM} (1977).
  \doi{10.1145/512950.512973}, \url{https://doi.org/10.1145/512950.512973}

\bibitem{DBLP:journals/jlp/CousotC92}
Cousot, P., Cousot, R.: Abstract interpretation and application to logic
  programs. J. Log. Program.  \textbf{13}(2{\&}3),  103--179 (1992).
  \doi{10.1016/0743-1066(92)90030-7},
  \url{https://doi.org/10.1016/0743-1066(92)90030-7}

\bibitem{d2021programmable}
D’Antoni, L., Hu, Q., Kim, J., Reps, T.: Programmable program synthesis. In:
  Computer Aided Verification: 33rd International Conference, CAV 2021, Virtual
  Event, July 20--23, 2021, Proceedings, Part I 33. pp. 84--109. Springer
  (2021)

\bibitem{DBLP:conf/cav/FedyukovichPMG19}
Fedyukovich, G., Prabhu, S., Madhukar, K., Gupta, A.: Quantified invariants via
  syntax-guided synthesis. In: Dillig, I., Tasiran, S. (eds.) Computer Aided
  Verification - 31st International Conference, {CAV} 2019, New York City, NY,
  USA, July 15-18, 2019, Proceedings, Part {I}. Lecture Notes in Computer
  Science, vol. 11561, pp. 259--277. Springer (2019).
  \doi{10.1007/978-3-030-25540-4\_14},
  \url{https://doi.org/10.1007/978-3-030-25540-4\_14}

\bibitem{gulwani2012synthesis}
Gulwani, S.: Synthesis from examples. In: WAMBSE (Workshop on Advances in
  Model-Based Software Engineering) Special Issue, Infosys Labs Briefings.
  vol.~10. Citeseer (2012)

\bibitem{10.1145/1993498.1993506}
Gulwani, S., Jha, S., Tiwari, A., Venkatesan, R.: Synthesis of loop-free
  programs. In: Proceedings of the 32nd ACM SIGPLAN Conference on Programming
  Language Design and Implementation. p. 62–73. PLDI '11, Association for
  Computing Machinery, New York, NY, USA (2011). \doi{10.1145/1993498.1993506},
  \url{https://doi.org/10.1145/1993498.1993506}

\bibitem{PGL-010}
Gulwani, S., Polozov, O., Singh, R.: Program synthesis. Foundations and
  Trends® in Programming Languages  \textbf{4}(1-2),  1--119 (2017).
  \doi{10.1561/2500000010}, \url{http://dx.doi.org/10.1561/2500000010}

\bibitem{qsygus}
Hu, Q., D'Antoni, L.: Syntax-guided synthesis with quantitative syntactic
  objectives. In: International Conference on Computer Aided Verification. pp.
  386--403. Springer (2018)

\bibitem{DBLP:conf/pldi/HuangQSW20}
Huang, K., Qiu, X., Shen, P., Wang, Y.: Reconciling enumerative and deductive
  program synthesis. In: Donaldson, A.F., Torlak, E. (eds.) Proceedings of the
  41st {ACM} {SIGPLAN} International Conference on Programming Language Design
  and Implementation, {PLDI} 2020, London, UK, June 15-20, 2020. pp.
  1159--1174. {ACM} (2020). \doi{10.1145/3385412.3386027},
  \url{https://doi.org/10.1145/3385412.3386027}

\bibitem{DBLP:conf/icse/JhaGST10}
Jha, S., Gulwani, S., Seshia, S.A., Tiwari, A.: Oracle-guided component-based
  program synthesis. In: Kramer, J., Bishop, J., Devanbu, P.T., Uchitel, S.
  (eds.) Proceedings of the 32nd {ACM/IEEE} International Conference on
  Software Engineering - Volume 1, {ICSE} 2010, Cape Town, South Africa, 1-8
  May 2010. pp. 215--224. {ACM} (2010). \doi{10.1145/1806799.1806833},
  \url{https://doi.org/10.1145/1806799.1806833}

\bibitem{kim2021semantics}
Kim, J., Hu, Q., D'Antoni, L., Reps, T.: Semantics-guided synthesis.
  Proceedings of the ACM on Programming Languages  \textbf{5}(POPL),  1--32
  (2021)

\bibitem{alpharegex}
Lee, M., So, S., Oh, H.: Synthesizing regular expressions from examples for
  introductory automata assignments. SIGPLAN Not.  \textbf{52}(3),  70–80
  (oct 2016). \doi{10.1145/3093335.2993244},
  \url{https://doi.org/10.1145/3093335.2993244}

\bibitem{10.1145/3632894}
Li, X., Zhou, X., Dong, R., Zhang, Y., Wang, X.: Efficient bottom-up synthesis
  for programs with local variables. Proc. ACM Program. Lang.  \textbf{8}(POPL)
  (jan 2024). \doi{10.1145/3632894}, \url{https://doi.org/10.1145/3632894}

\bibitem{DBLP:conf/pldi/MiltnerPMW20}
Miltner, A., Padhi, S., Millstein, T.D., Walker, D.: Data-driven inference of
  representation invariants. In: Donaldson, A.F., Torlak, E. (eds.) Proceedings
  of the 41st {ACM} {SIGPLAN} International Conference on Programming Language
  Design and Implementation, {PLDI} 2020, London, UK, June 15-20, 2020. pp.
  1--15. {ACM} (2020). \doi{10.1145/3385412.3385967},
  \url{https://doi.org/10.1145/3385412.3385967}

\bibitem{DBLP:conf/cav/NiemetzPRBT18}
Niemetz, A., Preiner, M., Reynolds, A., Barrett, C.W., Tinelli, C.: Solving
  quantified bit-vectors using invertibility conditions. In: Chockler, H.,
  Weissenbacher, G. (eds.) Computer Aided Verification - 30th International
  Conference, {CAV} 2018, Held as Part of the Federated Logic Conference, FloC
  2018, Oxford, UK, July 14-17, 2018, Proceedings, Part {II}. Lecture Notes in
  Computer Science, vol. 10982, pp. 236--255. Springer (2018).
  \doi{10.1007/978-3-319-96142-2\_16},
  \url{https://doi.org/10.1007/978-3-319-96142-2\_16}

\bibitem{DBLP:conf/sat/NotzliRBNPBT19}
N{\"{o}}tzli, A., Reynolds, A., Barbosa, H., Niemetz, A., Preiner, M., Barrett,
  C.W., Tinelli, C.: Syntax-guided rewrite rule enumeration for {SMT} solvers.
  In: Janota, M., Lynce, I. (eds.) Theory and Applications of Satisfiability
  Testing - {SAT} 2019 - 22nd International Conference, {SAT} 2019, Lisbon,
  Portugal, July 9-12, 2019, Proceedings. Lecture Notes in Computer Science,
  vol. 11628, pp. 279--297. Springer (2019).
  \doi{10.1007/978-3-030-24258-9\_20},
  \url{https://doi.org/10.1007/978-3-030-24258-9\_20}

\bibitem{DBLP:journals/corr/abs-2312-06001}
Padhi, S., Polgreen, E., Raghothaman, M., Reynolds, A., Udupa, A.: The sygus
  language standard version 2.1. CoRR  \textbf{abs/2312.06001} (2023).
  \doi{10.48550/ARXIV.2312.06001},
  \url{https://doi.org/10.48550/arXiv.2312.06001}

\bibitem{DBLP:conf/vmcai/PolgreenRS22}
Polgreen, E., Reynolds, A., Seshia, S.A.: Satisfiability and synthesis modulo
  oracles. In: Finkbeiner, B., Wies, T. (eds.) Verification, Model Checking,
  and Abstract Interpretation - 23rd International Conference, {VMCAI} 2022,
  Philadelphia, PA, USA, January 16-18, 2022, Proceedings. Lecture Notes in
  Computer Science, vol. 13182, pp. 263--284. Springer (2022).
  \doi{10.1007/978-3-030-94583-1\_13},
  \url{https://doi.org/10.1007/978-3-030-94583-1\_13}

\bibitem{polozov2015flashmeta}
Polozov, O., Gulwani, S.: Flashmeta: A framework for inductive program
  synthesis. In: Proceedings of the 2015 ACM SIGPLAN International Conference
  on Object-Oriented Programming, Systems, Languages, and Applications. pp.
  107--126 (2015)

\bibitem{DBLP:conf/cav/ReynoldsBNBT19}
Reynolds, A., Barbosa, H., N{\"{o}}tzli, A., Barrett, C.W., Tinelli, C.:
  cvc4sy: Smart and fast term enumeration for syntax-guided synthesis. In:
  Dillig, I., Tasiran, S. (eds.) Computer Aided Verification - 31st
  International Conference, {CAV} 2019, New York City, NY, USA, July 15-18,
  2019, Proceedings, Part {II}. Lecture Notes in Computer Science, vol. 11562,
  pp. 74--83. Springer (2019). \doi{10.1007/978-3-030-25543-5\_5},
  \url{https://doi.org/10.1007/978-3-030-25543-5\_5}

\bibitem{egg}
Willsey, M., Nandi, C., Wang, Y.R., Flatt, O., Tatlock, Z., Panchekha, P.: Egg:
  Fast and extensible equality saturation. Proc. ACM Program. Lang.
  \textbf{5}(POPL) (jan 2021). \doi{10.1145/3434304},
  \url{https://doi.org/10.1145/3434304}

\end{thebibliography}

\end{document}